# Security of Patched DNS

Amir Herzberg and Haya Shulman

Computer Science Department,
Bar Ilan University,
Ramat Gan, Israel
`{amir.herzberg,haya.shulman}@gmail.com`

**Abstract.** In spite of the availability of DNSSEC, which protects against cache poisoning even by MitM attackers, many caching DNS resolvers still rely for their security against poisoning on merely validating that DNS responses contain some 'unpredictable' values, copied from the request. These values include the 16 bit identifier field, and other fields, randomised and validated by different 'patches' to DNS. We investigate the prominent patches, and show how attackers can circumvent all of them, namely:
- We show how attackers can circumvent source port randomisation, in the (common) case where the resolver connects to the Internet via different NAT devices.
- We show how attackers can circumvent IP address randomisation, using some (standard-conforming) resolvers.
- We show how attackers can circumvent query randomisation, including both randomisation by prepending a random nonce and case randomisation (0x20 encoding).

We present countermeasures preventing our attacks; however, we believe that our attacks provide additional motivation for adoption of DNSSEC (or other MitM-secure defenses).

## 1 Introduction

Correct and efficient operation of the Domain Name System (DNS) is essential for the operation of the Internet. However, there is a long history of vulnerabilities and exploits related to DNS, mostly focusing on *DNS poisoning*. In a poisoning attempt the attacker causes recursive DNS servers (resolvers) to cache an incorrect, fake DNS record, e.g., mapping VIC-Bank.com to an IP address controlled by the attacker. DNS poisoning can facilitate many other attacks, such as injection of malware, phishing, website hijacking/defacing and denial of service.

The main technique for DNS poisoning is by sending forged responses to DNS requests which were sent by resolvers; to foil this, resolvers validate responses using different mechanisms. Currently, most resolvers rely only on non-cryptographic validation, mainly, validating that the response echoes some unpredictable (random) values sent with the request, such as in the DNS ID field, the client port, or within the resource (domain) name; e.g., see RFC 5452 [1]. Obviously, such mechanisms are insecure against a Man-in-the-Middle (MitM) attacker, who can read the randomness from the request and send a fake response with the valid identifiers.

Furthermore, even a weaker - and more common - *off-path, spoofing attackers*, may be able to send valid DNS responses and cause DNS poisoning, when the validated values are predictable or limited. For example, some DNS implementations use predictable identifiers (sequential, or using a weak pseudorandom generator); e.g., in [2], Klein shows how to predict the identifier for the then-current version of Bind 9, a widely-used DNS server, and how this can be exploited for highly-efficient DNS poisoning by a spoofing attacker. Indeed, as pointed out already in 1995 by Vixie [3], the identifier field alone is simply too short (16 bits) to provide sufficient defense against a determined spoofing attacker, who can foil it by sending many (but not too many) fake responses.

To improve DNS security, the IETF published DNSSEC [4,5,6], an extension to DNS, using cryptography (signatures and hashing) to ensure security (even) against MitM attackers. However, in spite of the publication of DNSSEC already in 1997 [7], and the wide awareness to its existence, deployment is still limited - e.g., less than 2% as reported in [8] for April, 2012. There are also many caching DNS resolvers that still do not support, or do not perform validation of, DNSSEC [9]; we discuss the deployment status of DNSSEC in Appendix B. Furthermore, due to implementation errors DNSSEC protection may fail, even when both the resolver and zone deploy it: validation of signatures of important top level domains, e.g., MIL, fails since the root does not delegate the public signature key of MIL but instead provides an incorrect indication that MIL does not support DNSSEC. This results in resolvers falling back to a non-validating mode.

Indeed the deployment of DNSSEC is progressing slowly, due to challenges (see [10]), and possibly due to the recent improvements ('patches') to non-cryptographic defenses, causing 'if it ain't broke, dont́ fix it' response. These patches are mainly by deploying new sources of 'unpredictability' in DNS requests and responses, such as use of random source ports [11,12], random DNS server selection [1] and random capitalisation of the domain name [13].

This manuscript focuses on relying on such non-cryptographic 'patches' to defend against DNS poisoning. This has two goals: to help improve these patches, since evidently they will remain widely used for years; and to further motivate adoption of more secure solutions such as DNSSEC, by pointing out weaknesses in the patches. While these specific weaknesses can be fixed (and we show how - often, easily), their existence should motivate the adoption of better security measures such as DNSSEC, providing security against MitM attacker and allowing for better validation of security, e.g., see [14].

### 1.1  Patching Caching DNS Resolvers Against Poisoning

Many researchers have identified vulnerabilities in the approach of relying on an 'unpredictability' of some fields in a DNS request and proposed patches;

we next review some of the main results. Bernstein, [15], suggested to improve DNS's defense against spoofed responses by sending the request from a *random port*, which can add a significant amount of entropy[1]. To prevent *birthday attack*, where attacker causes resolver to issue multiple queries for same domain in order to increase the probability of a match with one of multiple fake responses, Bernstein [15] and others suggest to limit the maximal number of concurrent requests for the same resource record (to one or to some small number); this technique is usually referred to as the *birthday protection*.

Many implementations did not implement these suggestions till the recent Kaminsky attack, [11,12], which introduced two critical improvements, allowing devastating attacks on many Internet applications. The first improvement was to control the time at which the resolver sends queries (to which the attacker wishes to respond), by sending to the resolver queries for a non-existing host name, e.g., with a random or sequential prefix of the domain name. The second improvement was to add, in the spoofed responses sent to the resolver, a type NS DNS record (specifying a new name for the domain name server) and/or a type A 'glue' DNS record (specifying the IP address of that domain's name server). These records poison the resolver's entries for the victim name server. Hence, if the attack succeeds once (for one record), the adversary controls the entire name space of the victim.

As a result of Kaminsky's attack, it became obvious that changes were needed to prevent DNS poisoning. Indeed, major DNS resolvers were quickly patched. The most basic patches were known measures - source port randomisation and birthday protection (see above).

These and other additional patches were summarised in RFC 5452 [1], including the use of random (valid) IP addresses for the name server. Additional patches, implemented by some resolvers, are to randomise DNS queries by randomly 'case toggling' the domain name (0x20 encoding [13]), or by adding a random prefix to the domain name [16]. It is tempting to interpret the analysis in [13,17,1,16] as indication that the 'patches' may suffice to make poisoning impractical, reducing the motivation for deployment of more systematic improvements. However, we caution against this conclusion. This work shows that in common scenarios, attackers can often circumvent some or all of the 'patches', making it still feasible to poison resolvers that rely on validation of 'unpredictable' values copied from requests to responses (rather than relying on cryptographic security, as in DNSSEC).

Some concerns with 'patches' were presented in earlier works. In particular, the most widely and easily deployed 'patch' is clearly source port randomisation. However, security experts, e.g., [18,12], noted that DNS resolvers located behind firewall/NAT devices, that use sequential assignment of external ports, were still vulnerable to the poisoning attack. On the other hand, it was widely believed that 'port-randomising' NAT devices, that sufficiently randomise the external ports, could retain or even improve the defense against DNS cache poisoning, e.g., see [12]. In addition, it was believed that 'port-preserving' NAT devices, that leave the source port intact (if it were not already allocated to another host), can be safely used with port-randomising resolvers, e.g., see [19]. Our results show otherwise, i.e., some of our attacks show how to circumvent port randomisation, in the resolver-behind-NAT scenario, even for port-randomising and port-preserving NATs.

---

[1] The exact amount of entropy added depends on the number of available ports, which may be below $2^{16}$.

Note that the resolver-behind-NAT scenario is common [20,21]. A recent study, [9], of DNSSEC deployment by recursive resolvers observed that a large number of recursive DNS resolvers is located behind NAT devices, and often many resolvers are even behind the same NAT device. Furthermore, [22] found that 90% out of 20,000 DSL lines (from a major European ISP) were located behind a NAT device.

### 1.2  Attacker Model

In our attacks, we assume an off-path, spoofing adversary connected to the Internet and a non-spoofing (user-mode) compromised host (zombie) on the local network; such zombies exist in many networks, e.g., see [23]; the attacker model is depicted in Figure 5, in the Appendix. The zombie can open user mode sockets and can send arbitrary (non-spoofed) packets, [24]. Alternately, the attacker can poison a DNS resolver which provides open recursion services.

Note that, in many scenarios, attackers do not even need to infiltrate malware to be able to control a host inside a local area network, e.g., attackers can obtain legitimate temporary access to organisational or other networks, or can connect to wireless networks as a legitimate client.

### 1.3  Our Contributions

The security of patched DNS resolvers relies on the randomness provided by the validation fields. We show that it is possible to reduce and often to nullify the randomness, thereby exposing the resolvers to Kaminsky style poisoning attacks. Our attacks apply to all widely-deployed 'patches':

SOURCE-PORT RANDOMISATION. We expose vulnerabilities in common source ports allocation algorithms, used by popular NAT devices; see Section 2. The vulnerabilities allow to predict the source port allocated to queries of the resolver. We tested our attacks in a lab setting against several NAT devices, see Table 1. The type of the NAT that the resolver resides behind is important in deciding which attack to launch.

DNS SERVER IP RANDOMISATION. We present techniques to predict (or force) the IP address of the name server to which the resolver will send its DNS request, see Section 3. Our techniques rely on fragmented DNS responses.

DOMAIN NAME RANDOMISATION. We further show that randomisation of DNS queries via 0x20 or by prepending a random string is not always effective and does not introduce protection against poisoning attacks; see Section 4.

In addition to presenting the vulnerabilities, we also propose countermeasures;

however, our most important contribution may be in motivating the adoption of systematic, secure defenses against poisoning, such as DNSSEC.

## 2  Source Port (de)Randomisation

In this section we show techniques to trap/predict the external port that will be allocated by the NAT device to the DNS request of the DNS resolver which the attacker wishes to poison. This phase allows to reduce (or nullify in some cases) the randomness added by source port randomisation of DNS resolvers. We tested popular NAT implementations, Table 1, that support different port allocation algorithms.

We identified the following common ports allocation algorithms: (1) *random allocation* (Section 2.1) where NAT selects ports at random from a pool of available ports until all ports are exhausted; (2) *per-destination sequential* (Section 2.2) where the NAT selects random port to each destination, and subsequent packets to that destination are allocated consecutive mappings; (3) *port preserving* (Appendix A), where NAT preserves the original port in the outgoing packet, and allocates sequentially upon collision; (4) *restricted random allocation*, where NAT maintains a mapping table that is smaller than the pool of available ports.

We also checked the source port allocation process of the NAT devices, which we tested in this work, via the DNS-OARC online *porttest* tool, [25]. The tool assigns one of the possible three scores: GREAT, GOOD and POOR, rating the 'unpredictability' of the ports allocation process. The tool reported a GREAT score for all NAT devices (except VMware NAT which allocates sequentially incremental ports). Yet, in this work we present techniques that allow the attacker to trap/predict the ports assigned to resolvers' DNS requests. The conclusion is that ports that 'appear' to be random should not be taken as indication of security. Indeed, as we show in this work, there are ways to circumvent this line of defense.

In what follows we show trap-then-poison (Section 2.1) and predict-then-poison (Section 2.2 and Appendix A) attacks for selected NAT devices; attacks for other NAT devices in Table 1 apply with slight variations and can be found in the extended version of this work [10]. Our descriptions and figures use illustrative choices of IP addresses, e.g., 10.6.6.6 (for zombie), 6.6.6.6 (for spoofing adversary, nicknamed Eve), 1.2.3.4 (for the authoritative name server of the 'victim' domain, V.com), and so on.

In all attacks, we exploit the fact that the NAT allocates mappings (permutations) between the addressing used by the internal host, identified by the 'pair' $< S_{IP} : S_{Port}, D_{IP} : D_{Port} >$, and the addressing used by the external host, identified by the 'pair' $< NATIP : NATPort, D_{IP} : D_{Port} >$, with the same values of $D_{IP} : D_{Port}$ in both pairs. We denote such mappings (permutations) by functions, e.g., $f(\cdot), g(\cdot)$. In our description of the attacks, we assume 'patched' DNS resolvers, that select unpredictable ('random') transaction ID and source port for each outgoing DNS request.

| Vendor | Port Allocation | Attack |
|---|---|---|
| Checkpoint (R70/FW-1) | restricted random (cannot be trapped) | resistant to attacks in Section 2 |
| Linux Netfilter Iptables (kernel 2.6) with '–random' | per-dest first random then sequential | Section 2.2 Predict attack |
| Linux Netfilter Iptables (kernel 2.6) | preserving (sequential if collide) | Section A Predict attack |
| VMware NAT (WS 6.5.2) | incremental | Kaminsky attack, [11] |
| Windows XP ICS (Service Pack 3) | first random then sequential | Section 2.2 Predict attack |
| Windows XP WinGate (Release 2.6.4) | preserving (random if collide) | Section 2 Trap attacks |
| CISCO IOS (release 15) | preserving (random if collide) | Section 2 Trap attacks |
| CISCO ASA (release 5500) | random (can be trapped) | Section 2 Trap attacks |

**Table 1.** Summary of the source port derandomisation attacks presented in this work, against different types of NAT devices that were tested.

Our attacks begin with a phase which allows the spoofer, Eve, to learn the port that will be allocated by the NAT to the DNS request of the DNS resolver. The port learning phase is performed with the help of a non-spoofing zombie, running user mode privileges.

### 2.1 Trap-then-Poison for Random Ports Allocation

The attack in this section relies on the fact that the NAT implements *outbound refresh mapping* for UDP connections, as specified in requirement 6 of RFC 4787 [26] (and implemented in most NATs). Namely, the NAT maintains the mappings from an internal (source) $S_{IP} : S_{port}$ pair to an external port $E_{port}$, for $T_{MR}$ seconds since a packet was last sent from $S_{IP} : S_{port}$ (in the internal side of the NAT) to the external network, using this mapping. We further assume that the NAT device selects an external port at random for each outgoing packet, e.g., CISCO ASA. The NAT device silently drops outgoing packets, sent from $S_{IP} : S_{port}$ to $D_{IP} : D_{port}$, when all external ports for $D_{IP} : D_{port}$ are currently mapped to other sources; this is the typical expected NAT behaviour [26].

The attack begins when the zombie contacts the attacker's command-and-control center, identifies its location, and receives signal to begin the attack. We next describe the steps of the attack; also illustrated with simplifications in Appendix, Figure 6.

**1.** The zombie, at address 10.6.6.6, sends UDP packets to 1.2.3.4:53, i.e., to the DNS port (53) of the name server of the 'victim' domain, whose FQDN is **ns.V.com**, from each of the $P$ ports in the set Ports. To handle faults, the payload of each packet contains the sending port $p \in$ Ports. Whenever the NAT receives one of these packets, it forwards it to **ns.V.com**, using some 'random' permutation $f$ over Ports; the allocation of each external port $f(p)$ to a specific internal port $p$ is held for $T_{MR}$ seconds, unless refreshed. Since none of these packets is a legitimate DNS packet, the authoritative name server ns.V.com ignores all of them, and does not send back any response.

**2.** After step 1 completed[2], Eve sends a packet with a spoofed source address 1.2.3.4:53, to external port 666 of the NAT (i.e., to 7.7.7.7:666). Since 7.7.7.7:666 is currently mapped to the internal IP address 10.6.6.6 and some port $f^{-1}(666)$, the NAT relays the packet to this IP and port. Thereby, the zombie learns the mapping of external port 666 to the internal port $f^{-1}(666)$; this will be crucial in the continuation of the attack, where we 'force' the query of the resolver to be sent using external port 666 (the 'trap'). This packet contains as a payload a random string of 8 or so digits to be used as the prefix of the FQDN in the query sent in the attack (in step 4).

**3.** After receipt of the packet on port $f^{-1}(666)$ in step 2, the zombie waits until the mappings established in step 1 are about to expire, i.e., until $t_3 = t_1 + T_{MR}$ (where $t_1$ is the time of step 1). At $t_3$, the zombie sends additional empty UDP packets, to all ports in Ports, *except* port $f^{-1}(666)$. As a result, the NAT refreshes the mappings on all of these ports; only the mapping for port 666 times out, and hence this becomes the *trap*: i.e., the only available external port of the NAT, which can be allocated for UDP packets whose destination is 1.2.3.4 : 53.

**4.** After step 3 completes, the attacker knows that the external port 666 of the NAT is the only port which can be allocated to the UDP packets sent from the internal network to the authoritative name server, at IP 1.2.3.4 : 53. The zombie sends a single DNS query to the resolver, for a random FQDN $r$.V.com; the use of a random 'subdomain' $r$ ensures that the resolver will have to issue a DNS query for this FQDN, since it surely will not have a cached value of the mapping. The resolver then sends a query to ns.V.com, from some 'random' (more precisely, unpredictable to attacker) port which we denote $p$, and using some random identifier $i$.

**5.** Next, at step 5, Eve sends a forged response per each $i \in 2^{16}$ values of the ID field. If one of these responses matches all of the validation fields in the query, the resolver will accept the poisoned records $r$.V.com A 6.6.6.6 and V.com NS $r$.V.com. Namely, from this point on, the resolver will consider 6.6.6.6 as a valid IP address for the authoritative DNS server of ns.V.com. The resolver will also forward the response $r$.V.com A 6.6.6.6 to the zombie, which will detect the successful attack, and inform Eve (this phase is not shown in the figure).

**6.** In step 6, the resolver receives the legitimate 'non-existing domain' (NXDOMAIN) response from the 'real' name server, at 1.2.3.4. If the attack succeeded, then this response is ignored, since the query is not pending any more. Otherwise, the resolver forwards the NXDOMAIN response to the zombie, who will inform Eve; they will repeat the attack from step 1 (as soon as ports expire on the NAT).

**7.** Finally, steps 7 and 8 illustrate subsequent poisoning of 'real' FQDN within the V.com domain. Since, following step 5, the resolver uses the 'poisoned' map-pings ns.V.com A 6.6.6.6, all subsequent requests for this domain are sent to 6.6.6.6.

---

[2] Eve can learn it is time to send the packet at the beginning of step 2, e.g., by an appropriate packet from the zombie to Eve upon completion of step 1.

### 2.2 Predict-then-Poison for Per-Destination Sequential Ports

In practice, due to efficiency considerations, NAT devices often do not select a random external port for *every* outgoing packet, but, depending on the NAT device, select the first port (for a tuple defined by $< S_{IP} : S_{Port}, D_{IP} : D_{Port}, protocol >$) at random, and subsequent ports are increased sequentially (for that tuple), until NAT refreshes its mapping for that tuple (if no packets arrived, e.g., after 30 seconds). For a different tuple, e.g., different destination IP, a new random port is selected for first packet, while subsequent packets are assigned sequentially increasing port numbers. When the NAT refreshes the mapping, i.e., by default 30 seconds, the port for outgoing packets with destination IP and port tuple is selected at random again. This behaviour is consistent with prominent NAT devices, e.g., Iptables NAT, Carrier Grade NAT [27].

In this section we present predict-then-poison attack on a *per-destination port randomising* NAT. A variation of the attack, which applies to *port preserving* NAT, appears in Appendix, Section A. In contrast to 'trap' attacks, the 'predict' attacks exploit an insufficient source port randomisation mechanism of the NAT, which allows to produce much more efficient attacks by predicting the source port allocated for the DNS requests by the NAT. In particular, the zombie is only required to generate and send three packets during the attack: first packet creates a mapping in the NAT table (so that packets from Eve can come through), subsequent packet lets Eve know which external port was used by the NAT, and the third packet is a DNS query which the zombie sends to local resolver for some random name in the victim domain $r$.V.com. The attack can be optimised by having the zombie transmit $k$ packets ($1 \leq k \leq 2^{16}$) from consecutive ports; Eve then sends $\lfloor \frac{P}{k} \rfloor$ packets, such that $j$-th packet is sent to port $j \cdot k$. Typically, it may be preferable for zombie to issue less packets (i.e., to use smaller $k$) to evade detection. The steps of the attack (in Figure 1) are described next.

**1.** Zombie opens the ports (to the destination IP address of the authoritative DNS). Specifically the zombie at address 10.6.6.6 sends $k$ UDP packets from sequentially increasing ports Ports[1],...,Ports[$k-1$]. All $k$ packets have 1.2.3.4:53 as the destination IP address and UDP port respectively (i.e., the name server of the victim domain, whose FQDN is **ns.V.com**). When the NAT receives the first packet (in the sequence of $k$ packets that were sent), it selects the first external port $x$ at random, the rest of $k-1$ packets are assigned consecutive (sequentially increasing) external ports. The NAT sends these $k$ packets to **ns.V.com**. The authoritative DNS **ns.V.com** ignores those UDP packets, since they do not constitute valid DNS requests.

**2.** Eve sends $\lfloor \frac{P}{k} \rfloor$ UDP packets, to sequentially increasing (by a factor of $k$) external ports of the NAT, with spoofed source IP 1.2.3.4:53. The payload of each packet contains the destination port number. The zombie will receive exactly one packet from Eve, w.l.o.g. on port $i^*$, and with payload containing $j^* \cdot k$ (i.e., packet that was sent to port $j^* \cdot k$ of the NAT).

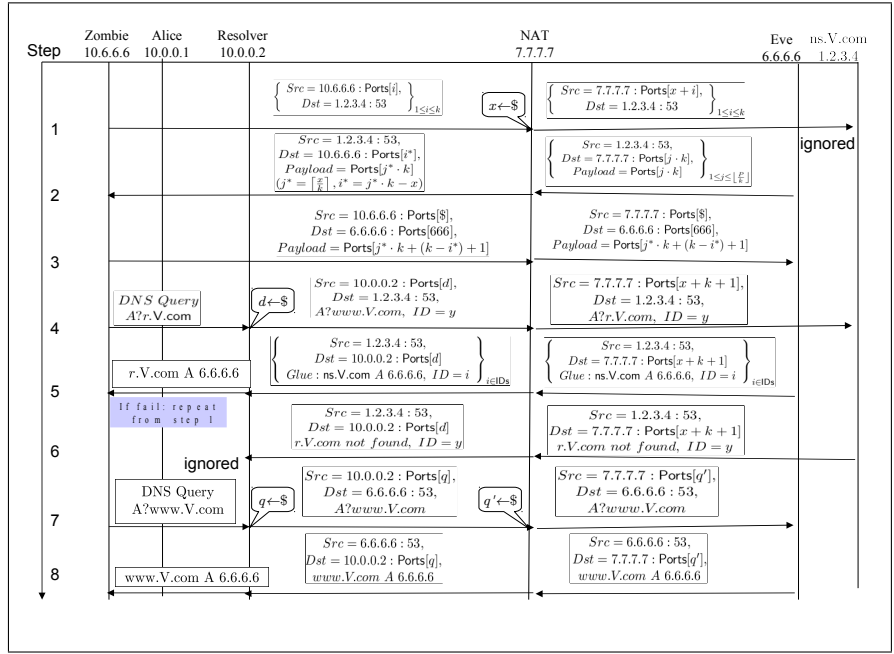

**Fig. 1.** Predict-then-poison DNS attack, for configuration in Figure 5, assuming per-destination port sequential NAT. The zombie sends a UDP packet to authoritative DNS ns.V.com, and queries the local resolver for the FQDN in the domain of the ns.V.com. Since the destination of the DNS request of the resolver is the same as the one used in the packet of the zombie, the NAT will not select the port at random but will assign a consecutive external port. This port can be predicted by Eve with the help of the zombie.

**3.** Next the zombie calculates the port that will be assigned by the NAT to the DNS query of the local resolver: $\mathsf{Ports}[j^* \cdot k + (k - i^*) + 1]$, and sends it to Eve in the payload (from some (random) source port $\mathsf{Ports}[\$]$ to a destination port 666, on which Eve is configured to be listening). Since the destination IP address of the packet sent to Eve is different from that of the authoritative name server, NAT will select an external port at random, and not consecutively, i.e., some $\mathsf{Ports}[\$]$ s.t., with high probability $\mathsf{Ports}[\$] \neq \mathsf{Ports}[x + k + 1]$.

**4.** The zombie then issues a DNS query to the local resolver, asking for a random FQDN $r.\mathbf{V.com}$. Since this domain name most likely does not exist in the cache, the resolver will send a DNS query from some (random) port $\mathsf{Ports}[d]$ containing a random identifier, to the authoritative name server **ns.V.com**. Note that the destination in the query of the local resolver is the same as the one that was used in the UDP packets of the zombie (i.e., the authoritative name server), the NAT will allocate the next available (consecutive) port, i.e., port $\mathsf{Ports}[x + k + 1]$, (to the query of the resolver), following the sequence of ports assigned to the packets of zombie.

**5.** As soon as Eve receives the packet containing the external port of the NAT that is mapped to the internal port of the resolver, she will generate and transmit $P$ packets with different values in the ID field, with spoofed source IP address (ostensibly originating from **ns.V.com**). The destination port in all the packets is $\mathsf{Ports}[j^* \cdot k + (k - i^*) + 1]$, and the answer field contains: $r.$**V.com** A 6.6.6.6 and **V.com** NS $r.$**V.com**. Since this port was allocated by the NAT to the query sent by the resolver, the NAT will forward all these DNS responses to the resolver.

**6.** Eventually when the authentic response 'non-existing domain' (NXDOMAIN) of the real name server at 1.2.3.4 arrives, the resolver will ignore it if one of the maliciously crafted packets (sent by Eve) matched and was accepted. The remaining steps are identical to the steps (7) and (8), presented in Section 2.1, Figure 6.

### 2.3   Experimental Evaluation

In this section we describe the setting that we used for validation of our attacks. We also summarise, in Table 1, our results for each NAT device against which we tested the attacks; the NAT devices were selected from different categories, i.e., proprietary NAT devices, e.g., Checkpoint, SOHO NAT devices, e.g., windows XP ICS, and other prominent NAT devices. This list of NAT devices that we tested is of course not exhaustive, but since we found that almost all of them, except one, allowed the attacker to reduce source port randomisation of the resolver, it is very likely that many more may be vulnerable to our (or other) attacks, e.g., Carrier Grade NAT of Juniper Networks (based on the technical report, [27], published in 2011).

TESTBED SETUP. Figure 5 illustrates the testbed used for the experimental evaluation of our attacks. The testbed consists of a NAT enabled gateway, which has two network cards. One card is connected via an ethernet cable to a switch, connecting a benign client, a compromised host, and a DNS resolver. The other is connected to Eve (also via a switch). The DNS resolver is running Unbound 1.4.1 software. The tests were run concurrently with other benign uses of the network, e.g., web browsing[3]. We used raw sockets to transmit the spoofed UDP packets sent by external spoofer Eve and user-level socket to transmit the UDP packets from zombie to Eve. We report on the results of the success of the DNS cache poisoning by running trap and predict attacks against several NAT devices in Table 1, and in more detail in the technical report [10].

### 2.4   Improved Port Allocation Mechanism

Official recommendations, [28], for NAT behaviour do not specify the instructions for the implementation of port allocation mechanism, as a result the developers and designers of NAT devices follow different approaches which may seem secure. Based on our findings we identify two design factors in ports allocation mechanism of the NAT: (1) the process via which the ports are selected (i.e., random, preserving, sequential); (2) the mapping table which maintains the allocated ports.

---

[3] Note that the TCP traffic, which constitutes the majority on the Internet, does not impact our attacks, since the NAT uses the PROTOCOL field in IP header when allocating external ports.

RANDOMISE PORTS SELECTION. Use port randomisation, but either with separate, random external port for each internal port, or at least with pseudo-random (but not sequential) increments between external port numbers[4]. Random ports assignment prevents the 'predict' attacks.

RESTRICTED MAPPING TABLE. The mapping table of allocated ports, maintained by the NAT, should be smaller than the pool of all the ports[5], e.g., half or less of the total of number of ports; a smaller mapping table prevents the attacker from trapping the port. For each arriving packet NAT should randomly select and assign a port from the pool of ports. Each time an entry is removed from the table when the external port is freed, e.g., the entry is refreshed after a timeout, NAT should select a new random port from the pool of ports.

## 3  IP Addresses (de)Randomisation

DNS resolvers can increase the entropy in DNS requests by randomising the IP addresses, i.e., selecting the source/destination IP addresses in the DNS request at random, and then validating the same addresses in the DNS response. Selecting random source IP address is rare, the resolvers are typically allocated one (or few) IP address as IPv4 addresses are a scarce resource. Furthermore, resolvers behind NAT devices use the IP of the NAT for their requests, and the address of the resolver is generally known [1].

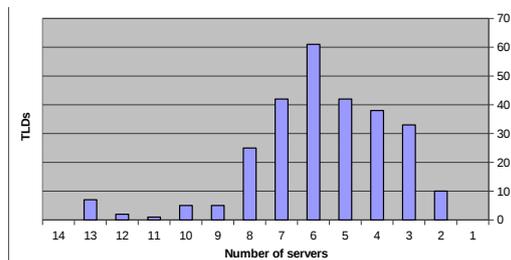

**Fig. 2.** The number of IP addresses in use by Top Level Domains (TLDs).

In contrast, most operators of DNS zones use a number of authoritative name servers for performance, robustness, and enhanced resilience to cache poisoning attacks. We found that the majority of top level domains (TLDs) use 5 to 7 authority name servers, and important domains, e.g., COM, use 13 authority servers[6]; see Figure 2.

When zone operators employ multiple authority servers, the resolver should send the query to the one with the shortest response time, and avoid querying non-responsive name servers, see [30,31]. However, there are no instructions on how to implement the server selection algorithm; as a result different resolvers, and even different versions thereof, implement different server selection algorithms, often resulting in inefficient implementations, [32].

---

[4] A pseudo-random permutation will provide as efficient data structure and lookup, as when using sequential allocation.

[5] This approach was supported only by the Checkpoint NAT which allowed it to evade our trap attacks.

[6] The list of TLDs is taken from the list published by IANA [29].

Indeed, the selection of the authority server by the caching resolver can often be predicted, e.g., if the attacker can measure the latency from the resolver to the authority name servers for a sufficient amount of time. However, this requires a significant effort from the attacker, and further may not result in precise prediction.

We focus on a weaker attacker which does not keep track of the latency to all the servers. However, our technique enables the attacker to predict the target name server's IP, for resolvers which avoid querying unresponsive name servers, as per the recommendations in [31,32]. We exploit the fact that when the target name server is not responsive, i.e., queries time-out, the resolver does not send subsequent queries to it, but only periodically, probes the target server until it becomes responsive. The (standard-conforming) Unbound (1.4.1) resolver sets this probing interval to 15 minutes. A similar behaviour was observed by [32] in PowerDNS, with the exception that PowerDNS sets the interval to 3 minutes. It appears that relying on the DNS server IP address randomisation for additional entropy requires careful study of particular resolver in question.

### 3.1   Predicting the Destination IP Address

The idea of destination IP prediction phase, in Figure 3, is to exploit large DNS responses which result in fragmentation; name servers that use TCP when responses exceed some predefined size, e.g., COM, are not susceptible to this attack. We performed the attack against a 404.GOV domain[7], whose *non-existing domain* responses exceed 1500 bytes and thus get fragmented en-route.

This phase, of forcing the resolver to use a specific IP, requires a puppet, i.e., a script confined in a browser, which issues DNS requests via the local caching DNS resolver, at IP 1.2.3.4 in Figure 3.

In steps 1 and 2 the puppet coordinates with the spoofer and issues a DNS request for $123.404.GOV (where $123 is a random prefix). In steps 3 and 4, the spoofer sends a forged second fragment, for all the possible name servers (i.e., a total of 2 spoofed fragments) except one which the attacker wants the resolver to use for its queries during the poisoning phase; the 404.GOV domain has three name servers. This ensures that only one IP address will result in a valid response, and the other two result in a malformed DNS packets. The spoofed second fragment is incorrect, and contains a single arbitrary byte (in addition to headers). In step 5, the spoofed second fragment is reconstructed with the authentic first fragment resulting in a malformed DNS packet which leaves the fragments reassembly buffer. This malformed DNS response is then discarded by the resolver, and the IP of the name server is marked[8] as 'non-responsive'.

---

[7] Many other zones return responses which get fragmented, e.g., MIL TLD; we focused on 404.GOV since it has only three name servers, which simplifies the presentation.

[8] In reality the resolver marks the server as 'non-responsive' after two unsuccessful respopnses, and this is easily handled by the attacker by sending two spoofed fragments with consecutive IP-ID in IP headers.

When the authentic second fragment arrives, it does not have a match and is discarded after a timeout. As a result the resolver does not receive the response, and after a timeout it resends the DNS request to the next DNS server, step 6. The same procedure applies here, and the response is discarded. In step 9 a valid response arrives from IP 162.138.183.11. In a real attack the resolver sends two queries to each server and marks the name server as non-responsive when *two* queries to that server result in a timeout; for simplicity in Figure 3 we present the process for one query to each server. This way, by wrecking the responses from all name servers except one, we forced the resolver to direct all its queries for 404.gov domain to 162.138.183.11.

Note that crafting a forged second fragment that would get matched with the authentic first fragment requires a match with the identification field (IP-ID) in the IP header. According to [33,34] the fragments of a datagram are associated with each other by their protocol number, the value in their IP-ID field, and by the source/destination address pair. Therefore the attacker is required to hit the correct IP-ID value, which is used by the name server in its DNS response. Many domains, as well as 404.gov, use per-destination sequential incrementing IP-ID values (or even globally sequential incrementing IP-ID, e.g., Windows OS). Other domains (mainly top level domains and the root servers) increment the IP-ID value in randomised quotas; we provide more details on this in [10].

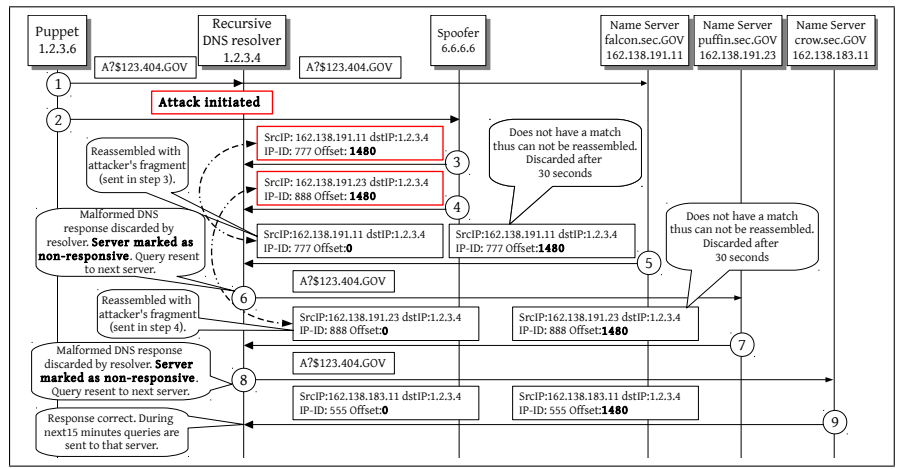

**Fig. 3.** The destination IP address prediction attacks: spoofing attacker crafts a forged second fragment that gets reassembled with the authentic first fragment and results in a malformed packet, which is discarded by the resolver.

The IP-ID allocation algorithm does not have a significant impact[9] on our attacks against Unbound (and alike) resolvers, as the number of 'misses', i.e., valid responses arriving to the resolver from some IP, does not prevent the attack and two failed (timed-out) queries suffice for Unbound to mark the server as non-responsive for 15 minute interval.

---

[9] Windows OS allows for a more efficient attack requiring less DNS queries.

### 3.2  Experimental Evaluation

The Wireshark capture, in Figure 4, that was run on the resolver, demonstrates the experimental evalutation, i.e., the DNS packets entering/leaving the network card of the resolver. During the course of the experiment the zombie issued 6000 queries[10] to the resolver. The spoofer initiates the attack by sending three spoofed fragments to each IP address except 162.138.183.11. For simplicity, the capture presents only the packets exchanged between the resolver and the name server of 404.gov at 162.138.191.23 (by adjusting a corresponding filter in wireshark); the complete capture contains queries/responses from other name servers too. Packets numbered 18-20 are the forged fragments sent by the spoofer, with sequentially incrementing IP-IDs. Then zombie triggers a DNS request, packet 29. The response from the name server contains two fragments, packets 33 and 34. The first fragment is reassembled with spoofed fragment 18, resulting in a malformed packet which is discarded by the resolver. The second fragment is discarded after a timeout. In packet 48 the resolver requests a public verification key of the 404.gov zone. The response contains three fragments 49-51; the first fragment is reconstructed with the spoofed fragment in packet 20, which also results in a malformed DNS response and is discarded. Note that this request, in packet 48, was sent at 19:28. Based on our tests it can be seen that when Unbound encounters a timeout twice for the same destination IP, it stops sending further packets to that destination for 15 minutes. Indeed, the next packet that is sent to that IP is packet number 6848, at time 19:43. The same scenario was observed with IP 162.138.191.11. The queries between 19:28 and 19:43 were sent only to 162.138.183.11, avoiding 162.138.191.11 and 162.183.191.23. Note that even if some of the responses (between packets 33 and 49) were valid and accepted by resolver, e.g., if they were not fragmented, it did not make a dif-ference, and two timed-out responses in a 15 minute interval were sufficient for Unbound to stop querying those IP addresses; this fact shows that the success probability of the attack does not depend on the IP-ID selection mechanism.

**Fig. 4.** The wireshark capture of the attack, presenting only the packets exchanged between the name server 162.183.191.23 and the resolver. As can be observed, after two malformed responses the resolver refrains from sending further queries to that name server for 15 minutes. Fragmented packets are coloured in white, DNS requests in black, and reassembled DNS fragments in blue.

---

[10] Note that our goal was to test the behaviour of the resolver, and to check the frequency of the queries to non-responsive servers; in real attack, once the IP-ID is known it is sufficient to issue two queries to mark the server as non-responsive.

### 3.3 Improved IP Address Randomisation

The attack we presented holds against a specific DNS resolver software, however we caution that variations of our ideas may apply to other server selection algorithms, and we believe that in the long term best answer to our derandomisation attacks is to deploy DNSSEC.

In the meanwhile we suggest (1) increasing the number of IP addresses, both of the name server and of the DNS resolver, e.g., an approach recently proposed by [35] is to superficially increase the number of IP addresses of the resolver for its DNS requests by *reusing* the IP addresses available to the network. Derandomising the IP addresses of the resolver seems to be a challenging task for the attacker; and (2) improving name server selection mechanisms, in particular, and contrary to the recommendations in [32], the server selection algorithms should not prefer less latent server and should not introduce differences in computation of RTT (round trip time).

## 4  DNS Query (de)Randomisation

In this section we describe two prominent defenses, 'case toggling' and random prefix, which are known to add significant extra entropy to DNS requests and show simple ways to circumvent them.

'cASe toGgLiNg'. Dagon et al. [13] present *0x20 encoding* for prevention of DNS poisoning. The technique is to randomly toggle the case of letters of which the domain name consists, and validate them in response. If the domain name $d$ contains $l(d)$ alphabetic characters, this increases the space of identifiers by a factor of $X(d) = 2^{l(d)}$. For example, a typical domain name, e.g., www.google.com, would increase the attacker's search space by a factor of $X(\text{www.google.com}) = 2^{12}$. Of course, the factor $X$ depends on the number of alphabetic characters in the domain name, e.g., $X(\text{a9.com}) = 2^4$. However, we believe that the distribution of domain queries with sufficient 0x20 characters, as reported by Dagon et al., is not indicative of the number of characters in queries that attackers will try to poison, and hence the impact of 0x20 encoding can be easily circumvented. In fact, in Kaminsky-style attacks, the query is intentionally for a non-existing domain name chosen by the attacker, e.g., the Top Level Domains (TLDs) such as .com and .uk; indeed the attackers prefer to poison a response to com rather than to www.google.com. Poisoning com allows the attacker a control over all subdomains of com (including www.google.com). Therefore, although 0x20 encoding provides significant extra entropy for long domain names, the attacker will choose to poison 'high value' TLDs, thereby circumventing the protection of 0x20.

RANDOM PREFIX. Prepending a random prefix to a DNS query[11] can ensure that a sufficiently large DNS query is sent, allowing to apply the 0x20 encoding on more letters and also making it more difficult for the attacker to guess the query (and the case of each letter).

The DNS query is composed of subdomains, at most 63 bytes each, separated by dots and at the right end the query ends with a top level domain, e.g., SUBDOMAIN.SUBDOMAIN.SUBDOMAIN.SUBDOMAIN.TLD. The total number of characters cannot exceed 255 bytes. So, prepending a random string $1 to query ABC.TLD, results in $1.ABC.TLD and increases the query by the size of $1.

But as we show this protection can also be foiled by the attacker. The attacker that wishes to poison an entry for some top level domain, e.g., COM, can issue a maximal size DNS query, i.e., 255 bytes, consisting of numbers, that will not allow prepending any more characters: 1-36.1-36.1-36.1-33.COM (the '1-36' denotes a string containing all numbers between 1 and 36). As a result, the attacker circumvents the 0x20 protection (which does not apply to numbers) and further avoids the addition of a random prefix to DNS request (since the query is already of maximal size). We tested the reduction of query randomisation (and foiling of 0x20) against Google Public DNS, [36], which supports both defenses, i.e., 0x20 and random prefix. A slight variation of this attack, see [10], also foils protection offered by WSEC DNS [16].

Such long queries to top level domains should be restricted or should receive a special treatment, yet we are not aware of appropriate recommendations in the relevant documentation.

## 5   Conclusions

Currently, most DNS resolvers defend against poisoning by validating that each response contains 'unpredictable' identifiers sent in the DNS query, such as the (16 bit) query ID. Such mechanisms are clearly insufficient to prevent poisoning by MitM attackers. A secure standard alternative exists: DNSSEC, which uses cryptography to achieve verifiable security. Indeed, DNSSEC deployment is progressing - but, quite slowly. Why so slowly? One reason are significant interoperability and performance concerns; another reason may be the existence of several 'patches', adding more 'unpredictable' identifiers. Such 'patches' are trivial to deploy and involve no or negligible overhead, hence, administrators may prefer to deploy them instead of deploying DNSSEC.

We study all major proposed 'patches', and find vulnerabilities in all of them. Our 'trap' and 'predict' attacks show that source ports may be disclosed or impacted by network devices such as NAT gateways. We show that the attacker can also nullify IP address randomisation of standard-conforming resolvers such as Unbound, forcing the resolver to query a specific name server. We also describe simple techniques to circumvent the DNS query randomisation via a random prefix and 0x20 encoding. We validated our attacks against popular NAT devices and standard DNS resolver software. Our derandomisation attacks are deployed 'sequentially' in phases, removing the randomisation of each identifier independently, and eventually strip the DNS request of the entropy offered by those 'unpredictability' fields, exposing the caching DNS resolvers to efficient poisoning attacks by off-path spoofing adversaries.

---

[11] A random prefix is a variation of the defense proposed in [16].

We show simple and effective countermeasures to our attacks. However, while using such 'patched patches' is tempting and easy, we believe that our work shows the importance of basing security on solid, strong foundations, as provided by DNSSEC, i.e., cryptographic protocols analysed to ensure security even against MitM attackers.

## A  Predict-then-Poison for Port-Preserving NAT

Port-preserving NAT leaves the original source port without modification when possible. When two clients (with different source IP addresses) send a packet (each) with the same source port, NAT preserves the port of the first packet and assigns the next available port to the second packet. As opposed to attack in Figure 1, the challenge here is to ensure that the query of the local resolver falls in the range of the ports occupied by the zombie. The coordination between Eve and zombie, when discovering the external port that will be used for the DNS query of the resolver is simpler, since Eve and zombie can coordinate in advance on the range size and initial and final indices of the packets sent by the zombie to the authority DNS. The steps of the attack are identical to the steps of the attack in Figure 1, with one exception: in step (2) Eve does not need to send $\frac{P}{k}$ packets and a single packet suffices. The number $k$ of UDP packets, that the zombie sends, should be sufficiently large, to increase the probability of the DNS packet of the local resolver to fall in the occupied range. The probability that the source port in a DNS query of the resolver will be in the range of ports occupied by the UDP packets of the zombie is $\frac{k}{P}$, where $k$ is the number of ports occupied by the zombie, and $P$ is the total number of ports.

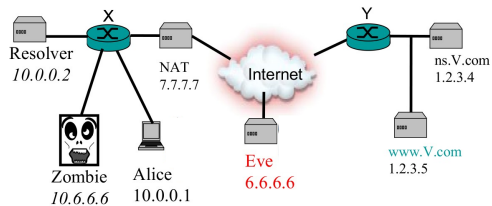

**Fig. 5.** The setting and the attacker model that we consider in all port-prediction attacks. DNS resolver is located behind a NAT device along with a benign client Alice and compromised machine Zombie. Spoofer Eve is located on the Internet and her goal is to spoof the DNS response for the victim domain $V.com$.

To evade detection[12] part of the attack can be carried out by the puppets, which can be used to make $k$ queries to the local resolver, to occupy sufficiently large range of ports, causing the query of the local resolver to be mapped to a predictable port.

---

[12] In order to maintain control over a zombie the attacker would prefer to reduce the usage of the zombie to minimum, and to use puppets when possible.

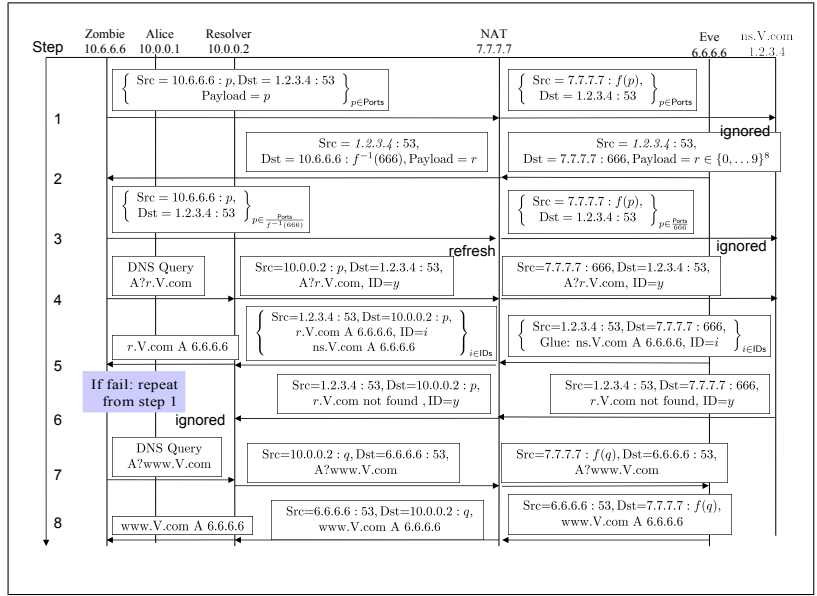

**Fig. 6.** DNS trap-then-poison attack using globally random ports allocation, for configuration in Figure 5.

## B  DNSSEC Deployment

DNSSEC ensures stronger security guarantees, however, its deployment remains to be seen due to the significant changes that it introduces to the current DNS infrastructure. In particular, DNSSEC requires cooperation and support by both the resolver and authoritative servers, which is a significant overhaul. DNSSEC introduces new resource records, e.g., signatures, which require specialised software for their generation, and also require changes to the zone file. Furthermore, there may be a reduced motivation to adop DNSSEC, as there are many non-conforming intermediate Internet devices, which discard fragmented traffic (DNSSEC enabled responses are large and often get fragmented), or large UDP packet. Many legacy proxy resolvers strip DNSSEC resource records from the DNS responses, rendering the DNS resolvers exposed to poisoning attacks.

Indeed recent studies [37,38,39,9] show that only about 0.02% of the zones have, so far, set up DNSSEC resource records; this indicates that the vast majority of organisations with an Internet presence are still vulnerable to attacks, e.g., [40]. A recent survey of *edu* names, [39], shows that about 1% are signed. The study's results also reveal that some fundamental capabilities required for adoption of DNSSEC in DNS resolvers, e.g., support of queries over TCP and support of EDNS0 [41], are not widely deployed.

The delay in deployment of the defenses against MitM adversaries, and the continuing attacks on DNS infrastructure, e.g., AT&T, Comcast and Rollingstone [42,43,44,40], motivates inspection of the 'easy-to-deploy' defenses, such as port randomisation.